\documentclass[reqno]{amsart} \usepackage{amscd}
\usepackage{epsf}
\newtheorem{theorem}{Theorem}[section]

\newtheorem{lemma}[theorem]{Lemma}

\theoremstyle{remark} 
\numberwithin{equation}{section}
\newcommand{\field}[1]{\ensuremath{\mathbb{#1}}}
\newcommand{\CC}{\field{C}}

\begin{document}
\title[Geometric quantization of self-duality ]{Geometric quantization 
of the moduli space of the Self-duality equations on a Riemann surface }
\author{Rukmini Dey}
\begin{abstract}
The self-duality equations on a Riemann surface arise as dimensional 
reduction of self-dual Yang-Mills equations. Hitchin had showed 
that the moduli space ${\mathcal M}$ of solutions of the self-duality 
equations on a compact Riemann surface of genus $g >1$ has a 
hyperK\"{a}hler structure. In particular ${\mathcal M}$ is a symplectic 
manifold. In this paper we elaborate on one of the symplectic structures, the 
details of which is missing in Hitchin's paper.  
Next we apply Quillen's determinant line bundle construction  
to show that ${\mathcal M}$ admits a prequantum line bundle. The Quillen 
curvature is shown to be proportional to the symplectic form mentioned 
above.  We do it in two ways, one of them is a bit unnatural (published in
R.O.M.P.) and a second way which is more natural. 
\end{abstract}

\maketitle


\section{Introduction}   
                                                          
Geometric prequantization is a construction of a Hilbert space, 
namely the square integrable  sections of a prequantum line bundle 
on a symplectic manifold $({\mathcal M}, \Omega)$ and  a correspondence 
between functions on ${\mathcal M}$
(the classical observables are functions on the phase space 
${\mathcal M}$) and operators on the Hilbert space such that the Poisson 
bracket of two functions corresponds to 
the commutator of the operators. The latter is ensured by the fact that 
the curvature of the prequantum line bundle is precisely the symplectic 
form $\Omega$ ~\cite{Wo}. 
Let $f \in C^{\infty}({\mathcal M})$. Let $X_f$  be the vector field 
defined by $\Omega(X_f, \cdot) = - df$. Let $\theta$ be the symplectic 
potential corresponding to $\Omega$. Then we can define the operator 
corresponding to the function $f$ to be 
$\hat{f} = -i h[X_f - \frac{i}{h}\theta(X_f)] + f$. 
Then if $f_1, f_2 \in C^{\infty}({\mathcal M})$ and 
$f_3 = \{ f_1, f_2 \}$, Poisson bracket of the two, 
then $[\hat{f}_1, \hat{f}_2] = -i h \hat{f}_3$. 

 A relevant example would be geometric
quantization of the  moduli space of flat connections. The moduli
space of flat connections of a principal $G$-bundle on a Riemann
surface has been quantized by  Witten 
by a construction of   the determinant line bundle of the Cauchy-Riemann 
operator, namely, ${\mathcal L}
= \wedge^{\rm{top}} ({\rm Ker} \bar{\partial}_A)^{*} \otimes \wedge^{\rm{top}}({\rm Coker}
\bar{\partial}_A)$, ~\cite{Wi}, ~\cite{ADW}. It carries 
the Quillen metric such that the canonical unitary connection  has a 
curvature form which coincides with the natural K\"{a}hler form on the 
moduli space of flat connections on vector bundles over the Riemann surface
 of a given rank ~\cite{Q}.

Inspired by ~\cite{ADW}, applying Quillen's determinant line bundle 
construction we construct  prequantum line bundles on the moduli space of 
solutions to the vortex equations ~\cite{D1} and the moduli space 
the self-duality equations over a Riemann surface which is a hyperK\"{a}hler 
manifold ~\cite{H}. In this paper we quantize one of the symplectic forms in two ways. In ~\cite{D} we show the quantization of the full hyperK\"{a}hler 
structure.

The self-duality equations on a Riemann surface arise from dimensional 
reduction of self-dual Yang-Mills equations from $4$ to $2$ 
dimensions ~\cite{H}.  
They have been  studied 
extensively in ~\cite{H}. They are as follows. Let $M$ be a compact Riemann 
surface of genus $g>1$ and let $P$ be a principal unitary
$U(n)$-bundle over $M$. Let $A$ be a unitary connection on $P$, 
i.e. $A = A^{(1,0)} + A^{(0,1)}$ such that 
$A^{(1,0)} = -A^{(0,1)*}$, where $*$ denotes conjugate transpose 
~\cite{GH},~\cite{K}. 
Let $\Phi$ be a complex Higgs field, 
$\Phi \in {\mathcal H} = \Omega^{1,0}(M; {\rm ad} P \otimes \CC)$.  
The pair $(A, \Phi)$ will be said to satisfy the self-duality equations if
$$(1)\rm{\;\;\;\;\;}F = -[\Phi, \Phi^*],$$
$$(2)\rm{\;\;\;\;\;}d^{\prime\prime}_A \Phi = 0.$$ 
Here $\Phi^* = \phi^* d \bar{z}$ where $\phi^*$ is taking 
conjugate transpose of the matrix of $\phi$.
Let the solution space to $(1)-(2)$ be denoted by $S$.
There is a gauge group acting on the space of $(A, \Phi)$ which leave the 
equations invariant. If $g$ is an $U(n)$ gauge transformation then 
$(A_1, \Phi_1)$ and $(A_2, \Phi_2)$ are gauge equivalent if 
$d_{A_2} g = g d_{A_1}$ and $\Phi_2 g = g \Phi_1$ ~\cite{H}, page 69.
  Taking the quotient by the gauge group of the solution 
space to $(1)$ and $(2)$ gives  the moduli space of solutions to these 
equations and is denoted by ${\mathcal M}$.
Hitchin shows that there is a natural metric on the moduli space 
${\mathcal M}$ and further  proves that the metric is 
hyperK\"{a}hler ~\cite{H}.

In the next section, we will elaborate on the natural metric and one of 
the symplectic forms, explicit mention of which is missing in ~\cite{H}. 
In the section after that we will construct a determinant line 
bundle on the moduli space ${\mathcal M}$ such that its Quillen 
curvature is precisely this symplectic form. 
This will put us in the context of geometric quantization.

We will do this in two different ways -- the first one using $\overline{\partial} + \overline{A_0^{(1,0)}} 
+ \overline{\Phi^{(1,0)}}$ which gauge transforms like  $\bar{g} (\overline{\partial} + \overline{A_0^{(1,0)}} 
+ \overline{\Phi^{(1,0)}}\bar{g}^{-1}.$ This is a bit unnatural since the 
action is with $\bar{g}$. (This approach was  published in 
R.O.M.P.)

In the next section we will do it using 
$\bar{\partial} + A_0^{(0,1)} + \Phi^{(0,1)}$
which gauge transforms like $g(\bar{\partial} + A_0^{(0,1)} + \Phi^{(0,1)})g^{-1}$ which is more natural.

In the end we discuss the holomorphicity of the prequantum line bundle
w.r..t the first complex structure.

In the paper where we construct prequantum line bundles for the full
hyperK\"{a}hler structure of the moduli space ~\cite{D} we 
mention the second
approach to quantizing the first symplectic form.

Papers which may be of interest in this context are ~\cite{BDr},~\cite{BDr1}, ~\cite{H1}, ~\cite{S} ~\cite{Wi1}. These papers use  algebraic geometry and 
algebraic topology and may provide alternative methods to quantizing the 
hyperK\"{a}hler system. Our method, in contrast,  is very elementary and we 
explictly construct 
the prequantum line bundles. The only machinery we use is Quillen's  
construction of the determinant line bundle, ~\cite{Q}.

Note: After writing the paper the author found  Kapustin and Witten's paper
~\cite{KW} where they have applied Beilinson and Drinfeld's quantization 
of the Hitchin system  to study the geometric 
Langlands programme.

\section{Symplectic structure of the moduli space}

This section is an elaboration of what is implicit in ~\cite{H}. 
Following the ideas in ~\cite{H} we give a
 proof that $\Omega$ below is a symplectic form on ${\mathcal M}$.

Let the configuration space be defined as ${\mathcal C} = \{ (A, \Phi)| 
A \in {\mathcal A}, \Phi \in {\mathcal H} \}$ where
${\mathcal A}$ is the space of unitary connections on $P$ 
and ${\mathcal H} = \Omega^{(1,0)}(M, {\rm ad} P \otimes \CC)$ is the space 
of Higgs field. Unitary connections satisfy $A = A^{(1,0)} + A^{(0,1)}$ 
where $A^{(1,0)*} = -A^{(0,1)}$, where $*$ is conjugate transpose. 
 
Let us define a metric on the complex configuration space
$$g ((\alpha, \gamma^{(1,0)}), (\beta, \delta^{(1,0)})) =  
- {\rm Tr} \int_M (   \alpha \wedge *_1 \beta)
-2 {\rm Im} {\rm Tr} \int_M  ( \gamma^{(1,0)} \wedge *_2 \delta^{(1,0){\rm tr}})$$ 
where 
$\alpha, \beta \in T_A {\mathcal A} = \{ \alpha \in \Omega^1(M, {\rm ad} P) |
 \alpha^{(1,0)*} = - \alpha^{(0,1)} \}$   
and $\gamma^{(1,0)},  \delta^{(1,0)} \in T_{\Phi}{\mathcal H} = 
\Omega^{1,0}(M; {\rm ad} P \otimes \CC).$ Here the superscript ${\rm tr} $ 
stands for  tranpose in the Lie algebra of $U(n)$, $*_1$ denotes 
the Hodge star taking $dx$ forms to $dy$ forms and $dy$ forms to $-dx$ forms 
(i.e. $*_1 (\eta dz) = -i \eta dz$ and 
$*_1(\eta d \bar{z}) = i \eta d \bar{z}$)  and 
 $*_2$ denotes the  operation (another Hodge star), 
 such that $*_2(\eta dz) = \bar{\eta} d \bar{z}$ and 
$*_2(\eta d \bar{z}) =  -\bar{\eta} d z$.

We check that this coincides with the   metric on the moduli space 
${\mathcal M}$ given by ~\cite{H}, page 79 and page 88. 
Hitchin identifies $A$ with its $A^{(0,1)}$ part. 
(Since it is a unitary connection, $A^{(1,0)}$ is 
determined by $A^{(0,1)}$ by the formula $A^{(1,0)} =- A^{(0,1)*}$). 
This is different from our point of view  where in  the connection 
part we keep both $A^{(1,0)}$ and $A^{(0,1)}$ even though they 
are related.  On $ T_{(A, \Phi)} {\mathcal C} 
= T_A {\mathcal A} \times T_{\Phi} {\mathcal H}$  
which is $\Omega^{(0,1)}(M, {\rm ad} P \otimes \CC) \times 
\Omega^{(1,0)}(M, {\rm ad} P \otimes \CC)$ for him, Hitchin defines 
a metric $g_1$ such that $g_1((\alpha^{(0,1)},\gamma^{(1,0)}), (\alpha^{(0,1)}, \gamma^{(1,0)})) 
= 2i {\rm Tr} \int_M (\alpha^{(0,1)*} \wedge \alpha^{(0,1)})
+ 2i {\rm Tr} \int_M (\gamma^{(1,0)} \wedge \gamma^{(1,0)*})$.   
$*$ denotes conjugate transpose as usual. 
Let $ \gamma^{(1,0)}= c dz$, where $c$ is a matrix. 

On $T_{(A, \Phi)} {\mathcal C}$, our metric 
\begin{eqnarray*}
& & g((\alpha, \gamma^{(1,0)}), (\alpha, \gamma^{(1,0)})) \\ 
&=&  - {\rm Tr} \int_M   (\alpha \wedge *_1 \alpha) 
- 2 {\rm Im} {\rm Tr} \int_M ( \gamma^{(1,0)} \wedge *_2 \gamma^{(1,0){\rm tr}})\\
&=& - {\rm Tr} \int_M (\alpha^{(1,0)} + \alpha^{(0,1)}) 
\wedge (-i \alpha^{(1,0)} + i \alpha^{(0,1)}) \\
& & -2 {\rm Im}  {\rm Tr} \int_M (c dz \wedge c^* d \bar{z}) \\
&=& 2i {\rm Tr} \int_M (\alpha^{(0,1)} \wedge \alpha^{(1,0)}) 
-2 Im    \int_M (-2i){\rm Tr} (c c^*) dx \wedge dy  \\
&=& -2i {\rm Tr} \int_M (\alpha^{(0,1)} \wedge \alpha^{(0,1)*}) 
+ 4 \int_M Re ({\rm Tr} (c c^*)) dx \wedge dy  \\
&=& 2i {\rm Tr} \int_M (\alpha^{(0,1)*} \wedge \alpha^{(0,1)}) 
+ 2i \int_M (-2i) {\rm Tr} (c c^*) .  dx \wedge dy \\
 &=& 2i {\rm Tr} \int_M (\alpha^{(0,1)*} \wedge \alpha^{(0,1)}) 
+ 2i \int_M {\rm Tr} (c c^*)  dz \wedge d \bar{z} \\
&=& 2i {\rm Tr} \int_M (\alpha^{(0,1)*} \wedge \alpha^{(0,1)}) 
+ 2i {\rm Tr} \int_M \gamma^{(1,0)} \wedge \gamma^{(1,0)*} \\
\end{eqnarray*} 
where we have used the fact that $\alpha^{(1,0)} = -\alpha^{(0,1)*}$ and that 
${\rm Tr} (cc^*)$ is real.  
Thus we get the same metric as Hitchin does.

The symmetry as $\alpha$ is interchanged with  $\beta$ and
$\gamma$ is interchanged with $\delta$ is as follows. In the first term 
$\alpha \wedge *_1 \beta = (\alpha^{(1,0)} + \alpha^{(0,1)}) 
\wedge ( *_1 \beta^{(1,0)} + *_1 \beta^{(0,1)}) 
= i (\alpha^{(1,0)} \wedge \beta^{(0,1)}) - i (\alpha^{(0,1)} 
\wedge \beta^{(1,0)})$. 
It is easy to check that $\beta \wedge *_1 \alpha$ is also exactly the same.  
In the second term, first note that 
${\rm Re} {\rm Tr} (AB^*) = {\rm Re} {\rm Tr} (BA^*)$ for matrices $A, B$. 
But we have matrix valued 
one forms.  Thus if $\gamma^{(1,0)} = A d z, \delta^{(1,0)} = B d z$, $A,B$ are matrices, 
then $*_2 \delta^{(1,0){\rm tr}} = B^* d \bar{z}$. Then the second term 
in the metric is the integral of
$-2 {\rm Im} {\rm Tr} (A dz \wedge B^* d \bar{z}) = 
-2{\rm Im} ({\rm Tr} (AB^*) dz \wedge d \bar{z}) = -2 {\rm Im} ({\rm Tr} (AB^*) . (-2i) dx \wedge dy) = -2 . {\rm Re} ({\rm Tr} (AB^*)). (-2) dx \wedge dy $  
Interchanging $\gamma$ with $\delta$ amounts to interchanging $A$ and $B$
which doesnot change the term.

There is an almost complex structure on ${\mathcal C}$, namely,
${\mathcal I} = \left[ \begin{array}{cc}
*_1  & 0 \\
0  & i
\end{array} \right]. $  Thus ${\mathcal I}(\beta, \delta^{(0,1)}) = 
(*_1 \beta, i \delta^{(0,1)})$. 

(If one identifies $T_{A} {\mathcal A}$ with $\Omega^{(0,1)} (M, {\rm ad }P \otimes \CC)$ as Hitchin does, then ${\mathcal I} = \left[ \begin{array}{cc}
i  & 0 \\
0  & i
\end{array} \right]$ i.e., ${\mathcal I}(\beta^{(0,1)}, \delta^{(0,1)}) = 
(i\beta^{(0,1)}, i \delta^{(0,1)})$. This is  the viewpoint we will take in a
paper sequel to this one. The metric and the symplectic forms will 
remain the same.)

We can define a symplectic form 
$$(3)\rm{\;\;\;}\Omega ((\alpha, \gamma^{(1,0)}), (\beta, \delta^{(1,0)})) 
= g ((\alpha, \gamma^{(1,0)}), {\mathcal I}(\beta, \delta^{(1,0)})) $$
Then 
\begin{eqnarray*}
 \Omega ((\alpha, \gamma^{(1,0)}), (\beta, \delta^{(1,0)})) 
&=&   {\rm Tr} \int_M (\alpha \wedge \beta)
-2 {\rm Im} {\rm Tr} \int_M (\gamma^{(1,0)} \wedge *_2(i\delta^{(1,0){\rm tr}})) \\
&=& {\rm Tr} \int_M (\alpha \wedge \beta)
-2  {\rm Im} {\rm Tr} \int_M ((-i) \gamma^{(1,0)} \wedge  *_2(\delta^{(1,0){\rm tr}})) \\
&=& {\rm Tr} \int_M (\alpha \wedge \beta)
+  2 {\rm Re} {\rm Tr} \int_M (\gamma^{(1,0)} \wedge \delta^{(1,0)*}) \\
&=& {\rm Tr} \int_M (\alpha \wedge \beta)
- {\rm Tr} \int_M (\gamma \wedge \delta)
\end{eqnarray*}
This is because $*_1^2=-1$,  $*_2(i \delta^{(1,0){\rm tr} }) 
= -i *_2 (\delta^{(1,0){\rm tr}})$, ${\rm Re} (-iz) = {\rm Im} z$ and 
finally 
$*_2(\delta^{(1,0){\rm tr}}) = \delta^{(1,0)*}$. Note that the first term of this 
symplectic form appears also in ~\cite{AB}, page 587. 

We will show by a moment map construction that this form descends to a 
symplectic form on the moduli space ${\mathcal M}$.

We need to find out the vector field generated by the action of the gauge 
group on ${\mathcal C}$ and hence on $S$ the solution space to $(1)-(2)$. 
Let $g = e^{ \epsilon \zeta}$, $\zeta \in u(n), $ the Lie algebra of the 
gauge group, i.e. $\zeta^* = - \zeta$.
Under the action of the gauge group connection 
$A \rightarrow A_g = gA g^{-1} + g dg^{-1} $ and 
$\Phi \rightarrow \Phi_g = g \Phi g^{-1} $. (The action on 
${\mathcal A}$  can be derived as follows: 
$(d + A_g) = g (d + A)g^{-1}$, by ~\cite{H}.
Thus, $(d + A_g) s =  g d(g^{-1}s)+ g A g^{-1} s 
= ds  + g d(g^{-1})s + g A g^{-1} s. $ 
Thus $A_g = g A g^{-1} + g d g^{-1}$ ).  
Taking $\epsilon$ to be very small,   we write $g = 1+ \epsilon \zeta$ and 
$ g^{-1} = 1 - \epsilon \zeta$ upto first order in $\epsilon$.   
Thus $A \rightarrow A - \epsilon ( d \zeta - [\zeta, A])$  
Similarly, $\Phi \rightarrow \Phi + \epsilon [ \zeta, \Phi] $. 
Thus on ${\mathcal C}$ there is a vector field generated by the 
gauge action given by 
$X_{\zeta} = (X_1, X_2) = (-( d \zeta - [\zeta, A]),[ \zeta, \Phi] )$. 

Define a moment map $\mu: {\mathcal C} \rightarrow u(n)^*$ as follows:
$$\mu(A, \Phi) =  (F(A) + [\Phi, \Phi^*]).$$ 
Given $\zeta$ as before, define the Hamiltonian to be 
$$ H_{\zeta} = {\rm Tr} \int_M (F(A) + [\Phi, \Phi^*])\zeta. $$
Now $F(A) = d A + A \wedge A$. 
Thus $F^{\prime} = lim_{t \rightarrow 0} \frac{F(A + t \beta ) - F(A)}{t} = d \beta + [\beta, A]$ where 
$\beta \in T_{A} {\mathcal A}$. (Note that for ${\rm ad} P$ 
valued $p$, $q$ forms $[\omega^{p}, \omega^{q}] = 
(-1)^{pq+1}[\omega^q, \omega^p]$, ~\cite{AB}, page 546.)
Let $$h_{\zeta} = {\rm Tr} \int_M F(A) \zeta = {\rm Tr} 
\int_M \zeta F(A)$$ (since ${\rm Tr} (AB)={\rm Tr} (BA)$). 
Now for $u, v, w \in \Omega^* (M, {\rm ad} P)$, 
$[u,v] \wedge w = u \wedge [v,w],$
~\cite{AB}, page 546. Thus $ {\rm Tr} (\zeta \wedge [\beta, A] ) 
= {\rm Tr} ([\zeta, \beta] \wedge A) 
= {\rm Tr} (-A \wedge [\zeta, \beta]) 
= {\rm Tr} (-[A, \zeta] \wedge \beta)
= {\rm Tr}([\zeta, A] \wedge \beta).$
Thus  if $\beta \in T_{A} {\mathcal A}= \Omega^{1}(M, {\rm ad} P),$   
\begin{eqnarray*}
d h_{\zeta}(\beta) &=& {\rm Tr} \int_M \zeta ( d\beta + [\beta, A])  \\
&=& {\rm Tr} \int_M (- d \zeta \wedge  \beta  + \zeta [\beta, A])\\
&=&  {\rm Tr} \int_M (-(d \zeta - [\zeta, A]) \wedge \beta)\\
&=&  {\rm Tr} \int_M (X_1 \wedge \beta). \rm{\;\;\;\;\;\;\;\;\;\;\;\;\;(4)}
\end{eqnarray*}

Let $$f_{\zeta}(\Phi) = {\rm Tr} \int_M ([\Phi, \Phi^*] \zeta),$$ 
and $\delta^{(1,0)} \in \Omega^{(1,0)}(M, {\rm ad} P \otimes {\CC}),$ 
the tangent space to ${\mathcal H}$.
Let $\delta^{(1,0)} = e dz$, $\Phi = \phi d z$, $\Phi^* = \phi^* d \bar{z}$,
$\delta^{(1,0)*}  = e^* d \bar{z}$, where $e$ and $\phi$ are matrices.

\begin{eqnarray*}
d f_{\zeta}(\delta^{(1,0)}) &=&  {\rm Tr} \int_M ([\delta^{(1,0)}, \Phi^*] \zeta 
+ [ \Phi, \delta^{(1,0)*}] \zeta)\\
&=& 2 {\rm Re}  {\rm Tr} \int_M ([\zeta, \Phi] \wedge \delta^{(1,0)*}) \\
&=& 2 {\rm Re}  {\rm Tr} \int_M (X_2 \wedge \delta^{(1,0)*}. \rm{\;\;\;\;\;\;\;\;\;(5)}
\end{eqnarray*}
This follows from the fact that 
\begin{eqnarray*}
2 {\rm Re} ({\rm Tr} [\zeta, \Phi] \wedge \delta^{(1,0)*}) &=& 2 {\rm Re} ({\rm Tr} ([\zeta, \phi] e^* )d z \wedge d {\bar z}) \\
&=& 2 {\rm Re} ({\rm Tr} [\zeta, \phi] e^*  (-2i d x \wedge dy )) \\
&=& 4 {\rm Im}  {\rm Tr} ([\zeta, \phi] e^*) d x \wedge dy 
\end{eqnarray*} 
Now,  
\begin{eqnarray*}
{\rm Im} {\rm Tr} ([\zeta , \phi] e^*) &=& 
{\rm Tr} ([\zeta,\phi] e^* - e [\zeta ,\phi]^*)/2i \\
&=& {\rm Tr}([\zeta ,\phi] e^* + e [\phi^*, \zeta])/2i \\ 
&=&  {\rm Tr}([\phi, e^*]\zeta + [e, \phi^*]\zeta)/2i.
\end{eqnarray*}
  Here we 
have used the fact that $\zeta^* = - \zeta$ since $\zeta \in u(n)$ and 
${\rm Tr} ([A,B]C)= {\rm Tr} ([B,C]A)$.
 
Thus, 
\begin{eqnarray*}
 4 {\rm Im}  {\rm Tr} ([\zeta, \phi] e^*) d x \wedge dy &=& {\rm Tr} ([\phi, e^*] \zeta + [e, \phi^*] \zeta) dz \wedge d \bar{z}\\
&=&  {\rm Tr} ([\Phi, \delta^{(1,0)*}] \zeta + [\delta^{(1,0)}, \Phi^*] \zeta)
\end{eqnarray*}

Thus $(5)$ follows.

From $(4)$ and $(5)$ it follows that
\begin{eqnarray*}
dH_{\zeta} ((\beta, \delta^{(1,0)})) &=& dh_{\zeta}(\beta) + df_{\zeta}(\delta^{(1,0)})\\
&=& {\rm Tr}  \int_M (X_1 \wedge \beta) 
+ 2 {\rm Re} {\rm Tr} \int_M (X_2 \wedge \delta^{(1,0)*}) \\
&=& \Omega(X_{\zeta}, (\beta, \delta^{(1,0)}))
\end{eqnarray*}
Therefore, 
 $$dH_{\zeta}(Y) = \Omega(X_{\zeta}, Y)$$
Thus the gauge group action on ${\mathcal C}$ is Hamiltonian 
and arises from a moment map.

\begin{lemma}
 $\Omega$ is a symplectic form on ${\mathcal M}$.  
\end{lemma}

\begin{proof}
We saw that equation $(1)$ is a moment map $\mu = 0$. 

About equation $(2)$  it is easy to check that ${\mathcal I}X$ 
satisfies the linearization of equation $(2)$ iff $X$ satisfies it. 
(This basically is as follows. Eq$(2)$ is 
$(\bar{\partial} + A^{(0,1)}) \Phi = 0$. Its linearization is 
$\bar{\partial} \gamma^{(1,0)} + \alpha^{(0,1)} 
\wedge \Phi + A^{(0,1)} \wedge \gamma^{(1,0)} = 0$, with 
$X= (\alpha, \gamma^{(1,0)}) \in T_{(A, \Phi)} {\mathcal C}$ as before. 
${\mathcal I} X = ((-i \alpha^{(1,0)}, i \alpha^{(0,1)}), i \gamma^{(1,0)})$. 
Since in ${\mathcal I} X$ both $\alpha^{(0,1)}$ and $\gamma^{(1,0)}$ 
just get multiplied by $i$, linearization of eq$(2)$ is satisfied by 
${\mathcal I} X$ iff it is satisfied by $X$). 

       If $S$ be the solution space to equation $(1)$ and $(2)$ and 
$X \in T_p S$ then ${\mathcal I}X \in T_p S$ iff $X \in T_{p}S$ is 
orthogonal to the gauge orbit $O_p = G \cdot p$. The reason is as follows. 
We let $X_{\zeta} \in T_p O_p,$ 
$g( X_{\zeta}, X) = \Omega (X_{\zeta}, {\mathcal I} X) = 
{\rm Tr} \int_M \zeta \cdot d \mu ({\mathcal I}X), $ and therefore 
${\mathcal I}X$ satisfies the linearization of equation $(1)$ iff 
$d \mu ({\mathcal I}X) = 0$ iff $g(X_{\zeta}, X) = 0$ for all $\zeta$.    
Similarly, $g(X_{\zeta}, {\mathcal I}X) = - \Omega(X_{\zeta}, X) 
= - {\rm Tr} \int_M \zeta d \mu(X) $.
Thus $X \in T_p S$ implies ${\mathcal I}X $ is orthogonal to $X_{\zeta}$ 
for all $\zeta$. Thus $X \in T_p S$ and $X$ orthogonal to $X_{\zeta}$ 
implies that ${\mathcal I}X \in T_p S$ and ${\mathcal I}X$ is orthogonal 
to $X_{\zeta}$.

Now we are ready to show that ${\mathcal M} $  has a natural
symplectic structure and an almost complex structure  compatible
with the symplectic form $\Omega $ and the metric $g$.

First we show that the almost complex structure descends to
 ${\mathcal M}$. Then using this and the symplectic quotient
construction we will show that $\Omega$ gives a symplectic
 structure on ${\mathcal M}$.
 To show that ${\mathcal I}$ descends as  an almost complex
 structure we let ${\rm pr}: S \rightarrow S/G =
 {\mathcal M}$ be the projection map and set $[p] = {\rm pr} (p)$. Then
we can naturally identify $T_{[p]}  {\mathcal M} $ with the
 quotient space $T_p S / T_p O_p, $ where $ O_p = G
 \cdot p $ is the gauge orbit. Using the metric $g$ on $S$ we can realize 
$T_{[p]} {\mathcal M}$ as a subspace in $T_p S$ orthogonal to  $T_p O_p$. 
Then by what is said before,
 this subspace is invariant under ${\mathcal I}$.
Thus $I_{[p]} ={\mathcal I} |_{T_p (O_p )^{\perp}}$, gives the
desired almost  complex structure. This construction does not
depend on the choice of $p$ since ${\mathcal I}$ is $G$-invariant.

The symplectic structure $\Omega$ descends to $\mu^{-1}(0) / G$,
(by what we said before  and by the Marsden-Wienstein symplectic
quotient construction ~\cite{GS}, ~\cite{H}) since the leaves of
the characteristic foliation are the gauge orbits. Now, as a
$2$-form $\Omega$ descends to ${\mathcal M}$ and so does the
metric $g$. We check that equation $(2)$ does not give
rise to new degeneracy of $\Omega$ (i.e. the only degeneracy of
$\Omega$ is due to $(1)$ but along gauge orbits). Thus $\Omega $
is symplectic on ${\mathcal M}$.
\end{proof}

\section{Prequantum line bundles}

A very clear description of
the determinant line bundle can be found in ~\cite{BF} and
~\cite{Q}. Here we mention  the formula for the Quillen curvature
of the determinant line bundle $\wedge^{{\rm top}}({\rm Ker} \bar{\partial}_A)^{*}
\otimes \wedge^{{\rm top}}({\rm Coker} \bar{\partial}_A) = {\rm det}(\bar{\partial}_A),$ where $\bar{\partial}_A = \bar{\partial} + A^{(0,1)}$ , 
given the canonical unitary
connection $\nabla_Q$, induced by the Quillen metric~\cite{Q}.
Namely, recall that the affine space ${\mathcal A}$ (notation as
in ~\cite{Q}) is an infinite-dimensional K\"{a}hler manifold. Here
each connection  is identified with its $(0,1)$ part. Since the
total connection is unitary (i.e. of the form $A= A^{(1,0)} + A^{(0,1)}$, 
where $A^{(1,0)} = -A^{(0,1)*}$) this identification is easy.
 In fact, for every $A \in {\mathcal A}$,
$T_A^{\prime} ({\mathcal A}) = \Omega^{0,1} (M, {\rm ad} P)$ and 
the corresponding K\"{a}hler form is given by 
$$F(\alpha, \beta) =  {\rm Re} {\rm Tr} \int_M  (\alpha^{(0,1)} \wedge  
\beta^{(0,1)*}) = - {\rm Re} {\rm Tr} \int_M  (\alpha^{(0,1)} \wedge  
\beta^{(1,0)}) $$   
where $\alpha^{(0,1)}, \beta^{(0,1)} \in \Omega^{0,1} (M, {\rm ad} P)$, and 
$\beta^{(1,0)} = - \beta^{(0,1)*}.$    
 It is skew symmetric 
if you interchange 
$\alpha^{(0,1)} = A d \bar{z}$ and $\beta^{(0,1)} = B d \bar{z}$ 
(follows from the fact that 
${\rm Im} ({\rm Tr} (A B^*)) = - {\rm Im} ({\rm Tr} (B A^*))$ for matrices $A$ 
and $B$, using once again $ d \bar{z} \wedge d z $ is imaginary).  
Let $\alpha = \alpha^{(0,1)} + \alpha^{(1,0)}$, $\beta =\beta^{(0,1)} + 
\beta^{(1,0)}$. 
It is clear from the fact that $ \alpha^{(1,0)} = -\alpha^{(0,1)*} $ and 
$\beta^{(1,0)} = -\beta^{(0,1)*}$ that $$ F(\alpha, \beta) = \frac{-1}{2} 
{\rm Tr} \int_M \alpha \wedge \beta. $$ 
Here we have used the fact that 
\begin{eqnarray*}
& & 2 {\rm Re} {\rm Tr} \int_M \alpha^{(0,1)} \wedge \beta^{(1,0)}\\
&=& {\rm Tr} \int_M \alpha^{(0,1)} \wedge \beta^{(1,0)} +  {\rm Tr} \int_M \overline{\alpha^{(0,1)}} \wedge \overline{\beta^{(1,0)}}\\  
&=& {\rm Tr} \int_M \alpha^{(0,1)} \wedge \beta^{(1,0)} + {\rm Tr} \int_M (-\overline{\alpha^{(0,1)\rm tr}}) \wedge (-\overline{\beta^{(1,0)\rm tr}}) \\
&=& {\rm Tr} \int_M \alpha^{(0,1)} \wedge \beta^{(1,0)} + {\rm Tr} \int_M \alpha^{(1,0)} \wedge \beta^{(0,1)} \\
&=& {\rm Tr} \int_M \alpha \wedge \beta 
\end{eqnarray*}

Then one has
$ {\mathcal F}(\nabla_Q) =  \frac{i}{\pi}  F. $

\subsection{Prequantization of the moduli space ${\mathcal M}$}

In this section we show that ${\mathcal M}$ admits a prequantum line bundle, 
i.e.a line bundle whose curvature is the symplectic form $\Omega$.
\begin{theorem}
The moduli space ${\mathcal M}$ of solutions to $(1)$ and $(2)$ admits a 
prequantum line bundle $P$ whose Quillen curvature 
${\mathcal F} = \frac{i}{\pi} \Omega$ where $\Omega$ is the natural 
symplectic form on ${\mathcal M}$ as in $(3)$.\end{theorem}

First we note that to the connection $A$ we can add any one form and
still obtain a derivative operator. To a  connection $A_0$ whose gauge 
equivalence class is fixed, we will add 
$\Phi$ to obtain new connections which will appear in the Cauchy-Riemann 
derivative operators. 

{\bf Definitions:} Let us denote by ${\mathcal L} = 
{\rm det} (\bar{\partial}+ A^{0,1})$ a determinant bundle on ${\mathcal A}$.

Let  ${\mathcal R} = {\rm det} (\overline{\partial} + \overline{A_0^{(1,0)}} 
+ \overline{\Phi^{(1,0)}})$  
where $A_0$ is a connection whose gauge equivalence class is fixed, 
i.e. $A_0$ is allowed to change only in the gauge direction. 
The ${\mathcal R}$ is a line bundle on ${\mathcal C}$ defined such that  
the fiber on $(A,\Phi)$ is that of 
${\rm det} (\overline{\partial} + \overline{A_0^{(1,0)}} 
+ \overline{\Phi^{(1,0)}})$, but the fiber over
gauge equivalent $(A_g, \Phi_g)$ is that of 
${\rm det}(\bar{g}(\overline{\partial} + \overline{A_0^{(1,0)}} 
+ \overline{\Phi^{(1,0)}})\bar{g}^{-1})$. This is because the gauge group  
${\mathcal G}$ acts on all of ${\mathcal A}$ simultaneously, so that when 
$A \rightarrow A_g$, $A_0 \rightarrow A_{0g}$.

(In the next section we have a better approach where we donot have to deal with $\bar{g}$).

Let ${\mathcal P} = {\mathcal L}^{-2} \otimes  
 {\mathcal R}^2$ denote a line bundle 
over ${\mathcal C}$. We will show that this line bundle is 
well defined on ${\mathcal M}$ and 
has Quillen curvature a constant multiple of the symplectic form $\Omega$.
 
\begin{lemma}
${\mathcal P}$ is a well-defined line bundle over 
${\mathcal M} \subset {\mathcal C}/{\mathcal G}$, 
where ${\mathcal G}$ is the gauge group.
\end{lemma}
\begin{proof}
First consider the Cauchy-Riemann operator 
$ D= \bar{\partial} + A^{(0,1)}$. Under gauge transformation 
$D=\bar{\partial} + A^{(0,1)}  
\rightarrow D_g= g(\bar{\partial} + A^{(0,1)})g^{-1} $.
We can show that the operators $D$ and $D_g$ have isomorphic 
kernel and cokernel and their corresponding Laplacians have the 
same spectrum and the eigenspaces are of the same dimension. Let 
$\Delta$ denote the Laplacian corresponding to $D$ and $\Delta_g$ 
that corresponding to $D_g$. Then $\Delta_g = g \Delta g^{-1}$. 
Thus the isomorphism of eigenspaces is  $s \rightarrow g s$. Thus when 
one identifies 
$\wedge^{{\rm top} }({\rm Ker} D)^* \otimes \wedge^{{\rm top} }
({\rm Coker} D)$ with 
 $\wedge^{{\rm top}}(K^a(\Delta))^* \otimes \wedge^{{\rm top}} 
(D(K^a(\Delta)))$ where $K^a(\Delta)$ is the direct sum of 
eigenspaces of the operator $\Delta$ of 
eigenvalues $< a$, over the open subset 
$U^a = \{A | a \notin {\rm Spec} \Delta \}$ of the affine space 
${\mathcal A}$ (see ~\cite{BF}, ~\cite{Q} for more details), 
there is an isomorphism of the fibers as $D \rightarrow D_g$. 
Thus one can identify 
$$ \wedge^{{\rm top}}(K^a(\Delta))^* \otimes \wedge^{{\rm top}} 
(D(K^{a}(\Delta))) \equiv
\wedge^{{\rm top}}(K^a(\Delta_g))^* \otimes \wedge^{{\rm top}} 
(D(K^{a}(\Delta_g))).$$
By extending this definition from 
$U^a$ to $V^a = \{(A, \Phi)| a \notin {\rm Spec} \Delta \}$, 
an open subset of ${\mathcal C}$,  we can define the fiber over 
the quotient space ${\mathcal C}/{\mathcal G}$ to be the 
equivalence class of this fiber. 

Similarly one can deal with the other case of 
${\rm det} (\overline{\partial} + \overline{A_0^{(1,0)}} + 
\overline{\Phi^{(1,0)}})$, 
because under gauge transformation, 
$\overline{\partial} + \overline{A_0^{(1,0)}} + 
\overline{\Phi^{(1,0)}} \rightarrow 
\bar{g}(\overline{\partial} + \overline{A_0^{(1,0)}} + 
\overline{\Phi^{(1,0)}})\bar{g}^{-1}$. 
Let $([A], [\Phi]) \in {\mathcal C}/{\mathcal G},$ 
where $[A], [\Phi]$ are gauge equivalence classes of $A, \Phi$, 
respectively.  Then associated to the equivalence class $([A], [\Phi])$ in the 
base space, there is an 
equivalence class of fibers coming from the identifications 
of ${\rm det}(\overline{\partial} + \overline{A_0^{(1,0)}} + 
\overline{\Phi^{(1,0)}})$ with 
${\rm det} (\bar{g}(\overline{\partial} + \overline{A_0^{(1,0)}} + 
\overline{\Phi^{(1,0)}})\bar{g}^{-1})$
as mentioned in the previous case. Note that in this case, the 
equivalence class of the fiber above $([A], [\Phi])$ is 
changing as $[\Phi]$ is changing. It is unaffected by change 
in $[A]$, since $[A_0]$ is  not changing. 

 This way one can prove that  ${\mathcal P}$ is well defined 
on ${\mathcal C}/{\mathcal G}$. Then we restrict it to 
${\mathcal M} \subset {\mathcal C}/{\mathcal G}$.
\end{proof}

{\bf Curvature and symplectic form:}

Recall $\alpha \in \Omega ^{1}(M, {\rm ad} P)$ has the decomposition 
$\alpha = \alpha^{(1,0)} + \alpha^{(0,1)}$, where 
$\alpha^{(1,0)}= - \alpha^{(0,1)*}$ . Similar decomposition holds 
for $\beta, \gamma, \delta \in \Omega^{1}(M, {\rm ad} P)$.  

Let $p = (A, \Phi) \in S$ where $S$ is the space of solutions to Hitchin 
equations $(1)$ and 
$(2)$. Let $X, Y \in T_{[p]}{\mathcal M}$. We write $X =(\alpha, \gamma)$ and $Y=(\beta, \delta)$, where 
$\alpha^{(0,1)}, \beta^{(0,1)} \in T_{A} ({\mathcal A}^{(0,1)}) = \Omega^{(0,1)}(M, {\rm ad} P \otimes \CC)$ and 
$\gamma^{(1,0)}, \delta^{(1,0)} \in T_{\Phi} {\mathcal H} = \Omega^{(1,0)}(M, 
{\rm ad} P \otimes \CC)$.
Since $T_{[p]}{\mathcal M}$ can be identified with a subspace in 
$T_p S$ orthogonal to $T_p O_p$ (the tangent space to the gauge orbit)  
 then 
$X,Y$ can be said to satisfy (a) $X, Y \in T_p S$ i.e. they satisfy 
linearization of $(1)$ and $(2)$ and (b) $X, Y$ are orthogonal to $T_p O_p $, the tangent space to the gauge orbit.

Let ${\mathcal F}_{{\mathcal L}^{-2}}$, $ {\mathcal F}_{{\mathcal R}^2}$,  
 denote the Quillen curvatures of the
determinant line bundles ${\mathcal L}^{-2}$, ${\mathcal R}^2$,  
 respectively.
Then, 
\begin{eqnarray*}
{\mathcal F}_{{\mathcal L}^{-2}}((\alpha, \gamma^{(1,0)}), (\beta, \delta^{(1,0)})) 
&=& -2 {\mathcal F}_{{\mathcal L}}((\alpha, \gamma^{(1,0)}), (\beta, \delta^{(1,0)}))\\ 
&=& - 2 \frac{i}{\pi} {\rm Re} {\rm Tr}\int_M (\alpha^{(0,1)}  
\wedge \beta^{(0,1)*}) \\
&=& \frac{i}{\pi} {\rm Tr} \int_M \alpha \wedge \beta
\end{eqnarray*}

\begin{eqnarray*}
{\mathcal F}_{{\mathcal R}^2}((\alpha, \gamma^{(1,0)}), 
(\beta, \delta^{(1,0)})) 
&=& 2{\mathcal F}_{{\mathcal R}}((\alpha, \gamma^{(1,0)}), 
(\beta, \delta^{(1,0)})) \\
&=& 2\frac{i}{\pi} {\rm Re} {\rm Tr} \int_M  \overline{\gamma^{(1,0)}} \wedge \overline{\delta^{(1,0)*}}\\
&=& 2\frac{i}{\pi} {\rm Re} {\rm Tr} \int_M  \gamma^{(1,0)} \wedge 
\delta^{(1,0)*} \\
&=& - \frac{i}{\pi}{\rm Tr} \int_M \gamma \wedge \delta
\end{eqnarray*}

Note:  $\overline{\gamma^{(1,0)}}$ and $\overline{\delta^{(1,0)*}}$ contributes because  of the term $\overline{\Phi^{(1,0)}}$ in ${\mathcal R}$. $\alpha$, $\beta$  donot contribute to this curvature  
because in the definition of ${\mathcal R}$ the gauge equivalence 
class of $A_0$ is fixed.

\begin{lemma} The Quillen curvature of ${\mathcal P} = {\mathcal L}^{-2} \otimes {\mathcal R}^2$ on ${\mathcal M}$ 
is $\frac{i}{\pi} \Omega$. 
\end{lemma}
\begin{proof} 
It is easy to check that
 ${\mathcal F}_{{\mathcal L}^{-2}}  
+ {\mathcal F}_{{\mathcal R}^2} 
= \frac{i}{\pi} \Omega$.
\end{proof}

These lemmas prove the theorem $(3.1)$.

\section{A more natural approach for the quantization}

We define $\Phi^{(0,1)}= -\Phi^{(1,0)*}$. We call the old $\Phi$ in the Hitchin
equations by $\Phi^{(1,0)}$.

 Let us denote by ${\mathcal L} = 
{\rm det} (\bar{\partial}+ A^{0,1})$ a determinant bundle on ${\mathcal A}$.

Let  ${\mathcal R} = {\rm det} (\bar{\partial} + A_0^{(0,1)} 
+ \Phi^{(0,1)})$  
where $A_0$ is a connection whose gauge equivalence class is fixed, 
i.e. $A_0$ is allowed to change only in the gauge direction.

Let ${\mathcal P} = {\mathcal L}^{-2} \otimes  
 {\mathcal R}^2$ denote a line bundle 
over ${\mathcal C} = {\mathcal A} \times {\mathcal H}$.

(This combination will give the prequantum line bundle corresponding to 
$\Omega$). 

\begin{lemma}
${\mathcal P}$ is a well-defined line bundle 
over ${\mathcal M} \subset {\mathcal C}/{\mathcal G}$, 
where ${\mathcal G}$ is the gauge group.
\end{lemma}
\begin{proof}
Let us consider the Cauchy-Riemann operator 
$ D= \bar{\partial} + A_0^{(0,1)} + \Phi^{(0,1)}$ which appears in 
${\mathcal R}$. The other case is analogous.
 Under gauge transformation 
$D=\bar{\partial} + A_0^{(0,1)} + \Phi^{(0,1)}  
\rightarrow D_g= g(\bar{\partial} + A_0^{(0,1)} + \Phi^{(0,1)})g^{-1} $ since it is the $(0,1)$ part of the connection operator $d + A_0 + \Phi$ which 
transforms in the same way.
We can show that the operators $D$ and $D_g$ have isomorphic 
kernel and cokernel and their corresponding Laplacians have the 
same spectrum and the eigenspaces are of the same dimension. Let 
$\Delta$ denote the Laplacian corresponding to $D$ and $\Delta_g$ 
that corresponding to $D_g$.The Laplacian is $\Delta = \tilde{D} D$ where 
$\tilde{D} = \partial + A_0^{(1,0)} + \Phi^{(1,0)}$, where recall $A_0^{(1,0)*} = -A_0^{(0,1)}$ and $\Phi^{(1,0)*} = - \Phi^{(0,1)}$. Note that  $\tilde{D} \rightarrow  \tilde{D}_g = g \tilde{D} g^{-1}$ under gauge transformation since it is 
the $(1,0)$ part of the connection operator $ d+ A_0+ \Phi$ which 
transforms in the same way.
 Thus $\Delta_g = g \Delta g^{-1}$.  
Thus the isomorphism of eigenspaces of $\Delta$ and $\Delta_g$ 
is  $s \rightarrow g s$. Thus when 
one identifies 
${\rm det} D$= $\wedge^{{\rm top} }({\rm Ker} D)^* \otimes \wedge^{{\rm top} }
({\rm Coker} D)$ with 
 $\wedge^{{\rm top}}(K^a(\Delta))^* \otimes \wedge^{{\rm top}} 
(D(K^a(\Delta)))$ where $K^a(\Delta)$ is the direct sum of 
eigenspaces of the operator $\Delta$ of 
eigenvalues $< a$, over the open subset 
$U^a = \{(A^{(0,1)} , \Phi^{(0,1)}) | a \notin {\rm Spec} \Delta \}$ of 
${\mathcal C}$ (see ~\cite{BF}, ~\cite{Q} for more details), 
there is an isomorphism of the fibers as $D \rightarrow D_g$. 
Thus one can identify 
$$ \wedge^{{\rm top}}(K^a(\Delta))^* \otimes \wedge^{{\rm top}} 
(D(K^{a}(\Delta))) \equiv
\wedge^{{\rm top}}(K^a(\Delta_g))^* \otimes \wedge^{{\rm top}} 
(D(K^{a}(\Delta_g))).$$  We can define the fiber over 
the quotient space $U^a/{\mathcal G}$ to be the 
equivalence class of this fiber. Covering ${\mathcal C}$ by open sets of the 
type $U^a$ enables us to define it on ${\mathcal C}/{\mathcal G}$. Then we 
restrict it to the moduli space ${\mathcal M} \subset 
{\mathcal C}/{\mathcal G}$.

 ${\mathcal L}$  also descends to the moduli space in the same spirit.
\end{proof}

{\bf Curvatures and symplectic forms}
 
Recall $\alpha \in \Omega ^{1}(M, {\rm ad} P)$ has the decomposition 
$\alpha = \alpha^{(1,0)} + \alpha^{(0,1)}$, where 
$\alpha^{(1,0)}= - \alpha^{(0,1)*}$ . Similar decomposition holds 
for $\beta, \gamma, \delta \in \Omega^{1}(M, {\rm ad} P)$.  

Let $p = (A, \Phi) \in S$ where $S$ is the space of solutions to Hitchin 
equations $(1)$ and 
$(2)$. Let $X, Y \in T_{[p]}{\mathcal M}$. We write $X =(\alpha, \gamma)$ and $Y=(\beta, \delta)$, where 
$\alpha^{(0,1)}, \beta^{(0,1)} \in T_{A} ({\mathcal A}^{(0,1)}) = \Omega^{(0,1)}(M, {\rm ad} P \otimes \CC)$ and 
$\gamma^{(1,0)}, \delta^{(1,0)} \in T_{\Phi} {\mathcal H} = \Omega^{(1,0)}(M, 
{\rm ad} P \otimes \CC)$.
Since $T_{[p]}{\mathcal M}$ can be identified with a subspace in 
$T_p S$ orthogonal to $T_p O_p$ (the tangent space to the gauge orbit)  
 then 
$X,Y$ can be said to satisfy (a) $X, Y \in T_p S$ i.e. they satisfy 
linearization of $(1)$ and $(2)$ and (b) $X, Y$ are orthogonal to $T_p O_p $, the tangent space to the gauge orbit.

Let ${\mathcal F}_{{\mathcal L}^{-2}}$, $ {\mathcal F}_{{\mathcal R}^2}$,  
 denote the Quillen curvatures of the
determinant line bundles ${\mathcal L}^{-2}$, ${\mathcal R}^2$,  
 respectively.
Then, 
\begin{eqnarray*}
{\mathcal F}_{{\mathcal L}^{-2}}((\alpha, \gamma), (\beta, \delta)) 
&=& -2 {\mathcal F}_{{\mathcal L}}((\alpha, \gamma), (\beta, \delta))\\ 
&=& - 2 \frac{i}{\pi} {\rm Re} {\rm Tr}\int_M (\alpha^{(0,1)}  
\wedge \beta^{(0,1)*}) \\
&=& \frac{i}{\pi} {\rm Tr} \int_M \alpha \wedge \beta
\end{eqnarray*}

(Since there is no $\Phi$-term in ${\mathcal L}$, 
$\gamma$ and $\delta$ donot contribute).

\begin{eqnarray*}
{\mathcal F}_{{\mathcal R}^2}((\alpha, \gamma), 
(\beta, \delta)) 
&=& 2{\mathcal F}_{{\mathcal R}}((\alpha, \gamma), 
(\beta, \delta)) \\
&=& 2\frac{i}{\pi} {\rm Re} {\rm Tr} \int_M  \gamma^{(0,1)} \wedge \delta^{(0,1)*}\\
&=& -2\frac{i}{\pi} {\rm Re} {\rm Tr} \int_M  \gamma^{(0,1)} \wedge \delta^{(1,0)}\\
&=& -2\frac{i}{\pi} {\rm Re} {\rm Tr} \int_M  \overline{(-\gamma^{(0,1)tr})} \wedge \overline{(-\delta^{(1,0)tr)})}\\
&=&-2\frac{i}{\pi} {\rm Re} {\rm Tr} \int_M  \gamma^{(1,0)} \wedge \delta^{(0,1)}\\
&=& 2\frac{i}{\pi} {\rm Re} {\rm Tr} \int_M  \gamma^{(1,0)} \wedge 
\delta^{(1,0)*} \\
&=& -\frac{i}{\pi}  {\rm Tr} \int_M  \gamma \wedge \delta
\end{eqnarray*}

Note:  $\gamma^{(0,1)}$ and $\delta^{(0,1)*}$ contributes because  of the term $\Phi^{(0,1)}$ in ${\mathcal R}$. $\alpha$, $\beta$  donot contribute to this curvature  
because in the definition of ${\mathcal R}$ the gauge equivalence 
class of $A_0$ is fixed. 

It is easy to check that the curvature of ${\mathcal P}$ is 
 $${\mathcal F}_{{\mathcal L}^{-2}}  
+ {\mathcal F}_{{\mathcal R}^2} 
= \frac{i}{\pi} \Omega.$$

Recall that $${\mathcal I} (\alpha^{(0,1)}) = i \alpha^{(0,1)},$$
$${\mathcal I}(\gamma^{(1,0)}) = i \gamma^{(1,0)},$$
$${\mathcal I}(\alpha^{(1,0)}) = -i \alpha^{(1,0)},$$
$${\mathcal I} (\gamma^{(0,1)}) = -i \gamma^{(0,1)}$$.

Thus w.r.t. ${\mathcal I},$  $A^{(0,1)}$ is holomorphic and $\Phi^{(0,1)}$ is antiholomorphic. 
But in ${\mathcal P}^{-1} = {\mathcal L}^2 \otimes {\mathcal R}^{-2}$ has the 
$A^{(0,1)}$-term as it is and the $\Phi^{(0,1)}$-term  in the inverse
bundle. Thus ${\mathcal P}^{-1}$ is ${\mathcal I}$-holomorphic.

Thus we have the following: 

\begin{theorem} ${\mathcal P}^{-1}$ is a holomorphic
 prequantum line bundle on ${\mathcal M}$ with curvature 
$-\frac{i}{\pi} \Omega$.
\end{theorem}

{\bf Polarization:} We can take ${\mathcal I}$-holomorphic sections of 
${\mathcal P}^{-1}$ as our Hilbert space.

School of Mathematics, Harish Chandra Research Institute, Jhusi,
Allahabad, 211019, India. email: rkmn@mri.ernet.in
\end{document}